# Ink-jet printing and drop-casting deposition of 2H-phase $SnSe_2$ and $WSe_2$ nanoflake assemblies for thermoelectric applications

B. Patil [1], C. Bernini [1], D. Marré [2,1], L. Pellegrino [1], I. Pallecchi [1]

[1] CNR-SPIN, Corso Perrone 24, 16152 Genova, Italy
[2] Università di Genova, Dipartimento di Fisica, Via Dodecaneso 33, 16146 Genova, Italy

## Abstract
The development of simple, scalable, and cost-effective methods to prepare Van der Waals materials for thermoelectric applications is a timely research field, whose potential and possibilities are still largely unexplored. In this work, we present a systematic study of ink-jet printing and drop-casting deposition of 2H-phase $SnSe_2$ and $WSe_2$ nanoflake assemblies, obtained by liquid phase exfoliation, and their characterization in terms of electronic and thermoelectric properties. The choice of optimal annealing temperature and time is crucial for preserving phase purity and stoichiometry and for removing dry residues of ink solvents at inter-flake boundaries, while maximizing the sintering of nanoflakes. An additional pressing is beneficial to improve nanoflake orientation and packing, thus enhancing electric conductivity. In nanoflake assemblies deposited by drop casting and pressed at 1 GPa, we obtained thermoelectric power factors at room temperature up to $2.2 \times 10^{-4}$ mW m$^{-1}$ K$^{-2}$ for $SnSe_2$ and up to $3.0 \times 10^{-4}$ mW m$^{-1}$ K$^{-2}$ for $WSe_2$.

## 1. Introduction

Graphene and related two-dimensional compounds represent a unique material platform for multifold technological applications, ranging from electronics, optoelectronics, spintronics to energy generation, energy storage, sensors. The common van der Waal structure of graphene and transition metal dichalcogenides (TMDs) not only makes them compatible for the fabrication of multifunctional heterostructures and composites, but it also allows to mutually borrow processing techniques and technological know-how, with minimal adaptation changes, yielding fast development and improvement of fabrication methods. Among low-cost large scale fabrication methods, liquid phase exfoliation (LPE) and subsequent deposition allows to obtain assemblies of nanoflakes with uniform distribution of size and thickness, deposited on any kind of substrates for specific applications, and possibly patterned for device design [1,2]. Besides these benefits, exfoliation of two-dimensional materials may result in improved properties as compared to bulk properties. For example, in a number of TMDs the structure can be changed when the thickness varies from bulk values to the atomically thin limit and the bandgap can be tailored by selecting the nanoflake thickness [3]. As another example, inter-flake boundaries in nanoflake assemblies may inhibit thermal transport, so that microstructure engineering can be carried out to improve thermoelectric performance [4,5]. Ink processing for thermoelectric materials has attracted attention, also thanks to simplicity and cost effectiveness of manufacturing [6].

LPE is a solution processing technique, where high shear forces generated in ultrasound sonication overcome the weak van der Waal forces and finally yield exfoliated flakes. The liquid medium should be a solvent with proper surface tension, solubility, and volatility properties. A pure solvent is preferable, as the amount of impurities and residues must be minimized in the exfoliated two-dimensional materials. However, binders can be added to improve particle aggregation and additives can be added to tailor wetting properties, rheological properties, surface tension and other functionalities. Subsequently, ink spreading and drying processes determine the morphology of the deposited assembly.

Deposition of exfoliated flakes by drop-casting is very simple, however obtaining a uniform coating and controlling the deposited thickness may be difficult by this method. On the other hand, inkjet printing has been the most used method in laboratory-scale demonstrators of printable two-dimensional material devices and applications, so far. Ink-jet printing is a non-contact, high-resolution (down to tens of μm), mask-less patterning technology, advantageous when minimal volume (1–2 mL) of low loading inks (e.g. LPE dispersions with <0.1 wt% of two-dimensional material concentration) are sufficient. It requires inks with low viscosity, 4-30 mPa s. If the challenges of maintaining stability in jetting, deposition and drying phases throughout an industrial process are addressed, this method could be scaled-up.

Ink-jet printed graphene with high electrical conductivity has been widely fabricated in view of different applications, namely printable and flexible electronics [7,8,9], optoelectronics [10], energy storage [11], and energy conversion [5] applications. Analogous processes of exfoliation and drop-casting or ink-jet deposition were also extended and adapted to other two-dimensional van der Waal materials, among which TMDs [9,12,13,14]. Biocompatibility was also addressed with the formulation of water-based and biocompatible inks of two dimensional materials [10,14,15,16] for biomedical applications.

Several literature reports demonstrated high potential of the LPE, drop-cast and ink-jet techniques to obtain conducting deposits of TMD nanoflake assemblies with good electric and thermoelectric properties for different applications. Electric properties are assessed in terms of high electrical conductivity σ, while assessment of thermoelectric properties is usually given by the figure of merit ZT, defined as $ZT = \frac{S^2\sigma}{\kappa}T$ where S, T, κ are respectively the Seebeck coefficient, the absolute temperature and the thermal conductivity. ZT represents the conversion efficiency between heat and electricity of thermoelectric materials. Another parameter of thermoelectric performance, which assesses only on the electronic properties, regardless phonon properties, is the thermoelectric power factor, which is the product of squared Seebeck coefficient and electrical conductivity, $S^2\sigma$. An intercalation, exfoliation and ink-jet printing was developed to obtain 5-10 micron-sized $MoS_2$ and $WSe_2$ flakes, with high exfoliation efficiency [17]. Ink-jet printed and drop-cast semiconducting $MoS_2$ were fabricated using different solvents and conductivity σ≈9x10$^{-2}$ Ω$^{-1}$ m$^{-1}$ was obtained [15]. Semiconducting $WS_2$ and $MoS_2$ films were obtained by liquid phase exfoliation plus drop-casting, ink-jet printing, or vacuum filtration [18,19]. Very recently, an ultra-fast process of ink-jet printing of aqueous precursors and rapid heating at 950°C-1050°C in reducing atmosphere was developed, allowing to obtain semiconducting TMDs films of hexagonal phase with centimeter size, controlled thickness, and high structural quality [14]. By this method, using domain merging on molten glass substrates and successive transfer, even single-domain monolayer $MoS_2$ and $MoSe_2$ of millimeter-size were obtained, having carrier mobilities as high as 21 and 54 cm$^2$ V$^{-1}$ s$^{-1}$ at room temperature, respectively [14]. A solution-based processing taking advantage multivalent cationic metal $Al^{3+}$ to bridge chemically exfoliated of semimetallic $TiS_2$ nanosheets was applied to obtain stable drop-cast flexible films [20]. Remarkably, in these $TiS_2$ samples electrical and thermoelectrical transport measurements indicated conductivity σ≈60000 Ω$^{-1}$ m$^{-1}$, S≈-60μV K$^{-1}$ and thus a power factor $S^2\sigma$≈217 μW K$^{-2}$m$^{-1}$. A process based on Li intercalation, exfoliation in water, pressing on PET substrates and annealing was used to obtain assemblies of semimetallic 1T-$MoS_2$ nanoflakes [21]. As compared to the more common and stable hexagonal n-type semiconducting 2H-$MoS_2$ phase, the tetragonal 1T-$MoS_2$ phase is p-type, semimetallic and 10$^7$ times more conducting than the 2H phase. In 1T-$MoS_2$ nanoflake assemblies, the noteworthy transport coefficients σ≈10000 Ω$^{-1}$ m$^{-1}$, S≈+85μV K$^{-1}$ and power factor S≈73 μW K$^{-2}$m$^{-1}$ were measured. Using electrospray and chemical exfoliation methods, crumpled $MoS_2$ nanoflakes with a prevalence of semimetallic 1T phase were obtained and they were ink-jet printed on silicon and photo-paper substrates, to fabricate electrodes for micrometric supercapacitors [22]. Thanks to conductivities σ≈2000 Ω$^{-1}$ m$^{-1}$ and aggregation-resistant properties, the supercapacitors with $MoS_2$ and reduced graphene oxide interdigitated electrodes exhibited wide stable-working voltage range and excellent capacitance retention. By using a process of chemical exfoliation, filtering and transfer from the filter membrane to a flexible substrate, films of semimetallic 1T phase of $WS_2$ and $NbSe_2$ with remarkable electric and thermoelectric properties were obtained, namely σ≈1000 Ω$^{-1}$ m$^{-1}$, S≈-75μV K$^{-1}$, and power factor $S^2\sigma$≈5 μW K$^{-2}$m$^{-1}$ for n-type $WS_2$ and σ≈150000 Ω$^{-1}$ m$^{-1}$, S≈+14μV K$^{-1}$, and power factor $S^2\sigma$≈35 μW K$^{-2}$m$^{-1}$ for p-type $NbSe_2$ [23]. An array of 100 stripes of such n-type $WS_2$ and p-type$NbSe_2$ was able to operate as a flexible thermoelectric generator, with 38 nW of output power generated by a temperature difference of 60K at room temperature, and exceptional stability to repeated mechanical bending and stretching [23].

In general, LPE processing of TMDs favors the formation of semimetallic 1T' phase to the detriment of semiconducting 2H phase [21,23]. On the other hand, annealing at temperatures higher than 50°C restores the more stable semiconducting 2H phase [21]. Much better results in terms of room temperature conductivity can be obtained with nanoflake assemblies of semimetallic 1T' phase and nanoflake composites containing semimetallic 1T' phase [20,21,23,24,25,26,27,28,29], yet nanoflake assemblies of semiconducting 2H phase could exhibit much higher Seebeck coefficient for optimal carrier concentration. The true potential of semiconducting 2H TMDs in the form of nanoflake assemblies in electric and thermoelectric applications has not yet been clarified.

In the above summary of literature results, it also emerges that obtaining good electric and thermoelectric transport properties in nanoflake assemblies obtained by LPE is challenging and was so far achieved on very few TMD compositions, namely $MoS_2$ and $WS_2$, mostly of the poorly stable 1T' semimetallic phase. On the other hand, the family of TMDs is extremely wide and could offer unexplored possibilities in terms of transport properties for device applications, especially in the semiconducting 2H phase, if combined with low-cost large scale fabrication methods.

In this work we selected 2H phase $SnSe_2$ and $WSe_2$, which are both semiconducting. Regarding $SnSe_2$, it is akin to the $SnS_2$ composition, where an inverse correlation between thermal and electrical conductivity was found with varying thickness, which is very interesting in view of optimizing electronic parameters for thermoelectric performance [30], and it is also akin to its monochalcogenide counterpart SnSe, whose thermoelectric properties are remarkable [31,32]. $SnSe_2$ itself was predicted by theory to reach thermoelectric figures of merit well above unity at high temperatures [33], and from experiments it was found to exhibit thermoelectric power factors up to 0.8 mW $K^{-2}$ $m^{-1}$ at room temperature, in the form of hot pressed nanopellets [34,35]. On the other hand, $WSe_2$ in the form of thin films and mechanically exfoliated flakes, exhibited very appealing thermoelectric properties in both electron-type and hole-type transport regimes, with power factors up to 4 mW $K^{-2}$ $m^{-1}$ at room temperature [36,37]. No less important, both these compounds have thermodynamic stability against oxidation or corruption, which allows flexibility in fabrication parameters and is crucial in view of thermoelectric applications.

In order to fill the literature gap on electric and thermoelectric transport properties of nanoflake assemblies obtained by LPE, which so far covers mostly the poorly stable 1T' semimetallic TMD phases, we prepared 2H phase $SnSe_2$ and $WSe_2$ nanoflake assemblies, by ink-jet printing or drop-casting. These compositions have never been prepared in the form of nanoflake assemblies obtained by LPE. Our choice of these compositions relied on the above cited literature reporting promising thermoelectric properties in $SnSe_2$ and $WSe_2$ single crystalline flakes, as well as in their thermodynamic stability. We carried out a systematic investigation of the relationship between preparation parameters (ink composition, deposition parameters, annealing temperature and time, uniaxial pressing) and electric and thermoelectric transport coefficients.

## 2. Methods

$WSe_2$- and $SnSe_2$-based inks with concentration 10 to 50 mg $mL^{-1}$ were purchased from BeDimensional [38]. Ethanol was chosen as solvent for its non-toxicity and its low boiling temperature. The suspended $WSe_2$ and $SnSe_2$ flakes had average lateral size 40 nm and 32 nm and average thickness 3.6 nm and 4.4 nm, respectively. Prior to drop casting or ink-jet printing, the inks were ultrasonicated for 15 min in an ice bath sonicator.

Liquid phase exfoliated nanoflake were deposited by drop casting and ink-jet printing on different substrates. Further additional processing steps were tried, namely annealing in reducing conditions and uniaxial pressing. Mica and quartz were chosen as substrates in the first stage of optimization and characterization, as they are rigid and easily handled, and also stable at high temperatures for thermal annealing tests. Clean mica surfaces were obtained by exfoliation and used as substrates; however, tendency to exfoliation made electrical contacts difficult to realize; therefore quartz substrates were used for further electric and thermoelectric characterization. Polyester and polyimide (kapton) were successively used as flexible substrates for pressing of deposits. Kapton was judged preferable with respect to polyester for mechanical reasons, namely lower flexibility. Annealing was performed at different temperatures (400°C-600°C) and times (2-3 hours), both in UHV and Argon atmosphere, to optimize removal of solvent residues and improve sintering without altering phase and stoichiometry. Uniaxial pressing was carried out at 1 GPa for 1 h.

In the case of drop-casting, multiple deposition steps were tried, and a number of around 30 drops was chosen as optimal compromise between amount of deposited material and homogeneity. The substrate was heated at 65 °C on hot plate during drop casting.

In the case of ink-jet printing, a Dimatix DMP-2831 Fujifilm ink-jet printer with 10 pL cartridge was used for deposition on mica and kapton substrates. The waveform and the printing parameters were adjusted by monitoring the drop characteristics through a dropwatch camera (see supplementary material for a video of the ejected drop and the used waveform for the printer, Fig. S1). For ink-jet printing of $WSe_2$ on kapton, the printing parameters were voltage around 16.9 V, maximum jetting frequency 5 kHz, drop spacing 10 µm, number of stacked layers 500, with pause every 40 layers to monitor the drop characteristics. The cartridge temperature was 28 °C and the substrate temperature during printing was 33°C. Printing was performed by using only one selected nozzle of the cartridge. Ink-jet printing of $WSe_2$ was also carried out on mica and kapton substrates, by depositing submillimetric patterns of successive lines, with drop spacing of 50 micron, line spacing 0.05 mm, number of layers 20, and time interval between two layers of 300 s.

Microstructure of samples was inspected by optical microscopy and Scanning Electron Microscopy (SEM). Stoichiometry was measured by Energy-dispersive X-ray spectroscopy (EDS), and acquired data were averaged on 8-10 different spots of the samples. Phase analysis was performed by Raman spectroscopy and X-rays diffraction. Resistance and thermopower measurements from room temperature down to low temperature were carried out in a commercial apparatus, a PPMS by Quantum Design, with home made adaptations, so as to carry out transport measurement of highly resistive samples. A four-probe geometry was used for electrical contacts, with evaporated gold electrodes. Thermopower measurements were performed in high vacuum and a steady-state technique was used to feed heat into the free-standing sample and establish a temperature gradient. A heater was attached at one sample edge, while the thermal sink was connected to the opposite edge. A sinusoidal current was applied to the heater and the output signals were extracted by Fast Fourier Transform of the voltage signals measured at voltage leads and thermocouples, respectively.

## 3. Results and discussion

### 3.1. Microstructural analysis of as-deposited samples

Images of $WSe_2$ nanoflake assemblies drop-cast on mica are shown in Fig. 1, taken either by optical (left) or electronic (right) microscopes. Granularity on a submicron scale and slight inhomogeneity on a scale of several tens of microns is visible. Optical and electronic images of ink-jet printed $WSe_2$ nanoflake assemblies on mica are shown in Fig. 2, (left and right, respectively). Granularity on a submicron scale is seen. The granularity of the $WSe_2$ flakes deposited by either drop-casting or ink-jet printer was similar, but in the case of drop casting the coverage was higher and the $WSe_2$ layer was denser.

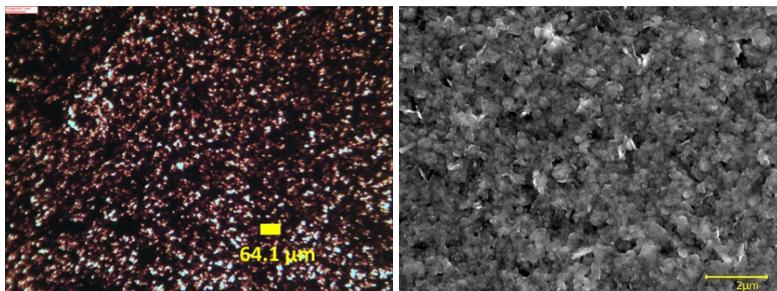

**Fig. 1:** Optical (left) and SEM Secondary Electron (right) images of $WSe_2$ nanoflake assemblies, deposited by drop-casting on mica and not annealed.

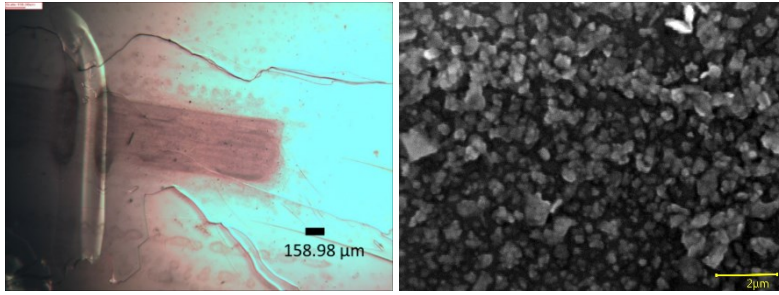

**Fig. 2:** Optical (left) and SEM Secondary Electron (right) images of WSe$_2$ nanoflake assemblies, deposited by ink-jet printing on mica and not annealed. The dark rectangle in the optical image is a pattern realized by multiple adjacent line scans of the ink-jet printer.

### 3.2. Phase and chemical analyses of drop-cast samples annealed at different temperatures and for different times

In order to obtain as high packing of nanoflakes as possible, tests of high temperature annealing of WSe$_2$ and SnSe$_2$ nano-flake assemblies deposited by drop-casting on quartz substrates were carried out. Samples of WSe$_2$ and SnSe$_2$ were annealed at different temperatures from 400°C to 800°C in a tube furnace in Ar flux and for different times from 1 h to 6 h. The obtained samples were characterized in terms of phase by x-rays diffractometry and in terms of stoichiometry by EDS. Two-probe resistance at room temperature was also measured. Table I summarizes the results on WSe$_2$ samples. X-rays patterns of these samples are reported in Fig. 3 (left) and resistance and stoichiometry values are plotted in Fig. 3 (right). With increasing annealing temperature, selenium evaporated and tungsten remained on the film surface, eventually transforming into conducting tungsten oxide and causing a significant drop of resistance in highly off-stoichiometric samples. On the other hand, annealing at 400°C preserved the nominal stoichiometry, resulting in samples with measured resistances in the MOhm range.

**Table I:** summary of the results of annealing tests carried out on WSe$_2$ drop-cast nanoflake assemblies.

| WSe$_2$ sample | Annealing details | Two-probe resistance | W:Se ratio | X-rays result | Remarks |
|---|---|---|---|---|---|
| WSe$_2$ on drop-cast quartz | Not annealed | Above measuring limit | ~ 0.5 | Correct phase | |
| WSe$_2$ on drop-cast quartz | @ 400 °C with 0.1 °C/s ramp and hold for 1 h | ~6-7 MOhm | ~ 0.59 | Correct phase | |
| WSe$_2$ on drop-cast quartz | @ 400 °C with 0.1 °C/s ramp and hold for 6 h | ~1.5 MOhm | ~ 0.66 | Correct phase | Cracks |
| WSe$_2$ on drop-cast quartz | @ 500 °C with 0.1 °C/s ramp and hold for 1 h | ~0.350 MOhm | ~ 0.81 | Correct phase, but WO$_3$ peaks present | |
| WSe$_2$ on drop-cast quartz | @ 600 °C with 0.1 °C/s ramp and hold for 1 h | ~0.060 MOhm | ~ 1 | Correct phase, but WO$_3$ peaks present | |
| WSe$_2$ on drop-cast quartz | @ 800 °C with 0.1 °C/s ramp and hold for 1 h | ~0.0013 MOhm | ~ 2.5 | Correct phase, but WO$_3$ peaks present | |

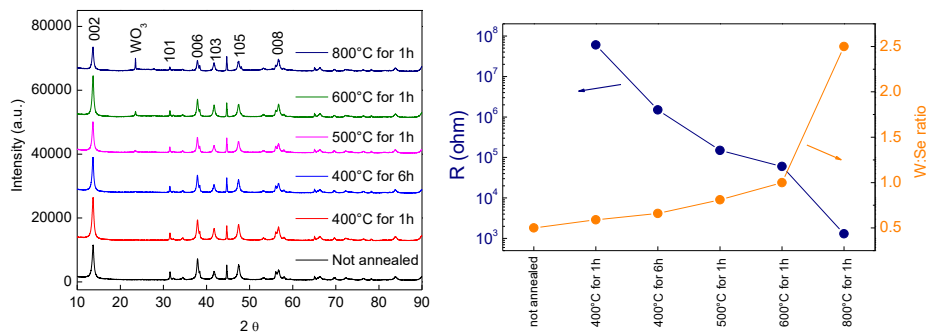

**Fig. 3:** Left: X-rays patterns of WSe$_2$ samples annealed at different temperatures and for different times. The Bragg indices of the WSe$_2$ phase are indicated. Right: trend of resistance and W:Se ratio in WSe$_2$ samples annealed at different temperatures and for different times.

Similar results were obtained in SnSe$_2$ samples (see Fig.4 and Table II). In this case, the decrease of resistance with increasing annealing temperature was related to off-stoichiometry due to evaporation of Se and presence of tin oxides. For annealing temperatures below 400°C, SnSe$_2$ assemblies were a mixture of SnSe$_2$ powder and nanoflakes. The Sn:Se ratio of the nano-flakes was 1:2, while the Sn:Se ratio of the powdery parts was approximately 2:3 (Sn:Se), indicating more tin in the sample, due either to segregated Sn or evaporated Se. A further difficulty was encountered with SnSe$_2$ samples, compared to WSe$_2$ samples, related to the powdery nature of the samples after annealing, which caused detachment from the quartz substrate.

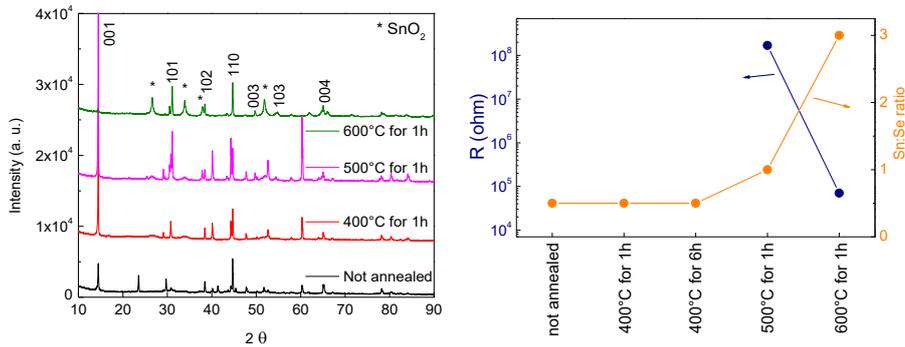

**Fig. 4:** Left: X-rays patterns of SnSe$_2$ samples annealed at different temperatures and for different times. The Bragg indices of the SnSe$_2$ phase are indicated. Right: trend of resistance and Sn:Se ratio in SnSe$_2$ samples annealed at different temperatures and for different times.

**Table II:** summary of the results of annealing test carried out on SnSe$_2$ drop-cast nanoflake assemblies.

| SnSe$_2$ sample | Annealing details | Two-probe resistance | SEM Sn:Se | XRD | Remarks |
|---|---|---|---|---|---|
| SnSe$_2$ drop-cast on quartz | Not annealed | Above measuring limit | ~0.5 | Correct phase | Sample is not so powdery and sticks to quartz |
| SnSe$_2$ drop-cast on quartz | @ 400 °C with 0.1 °C/s ramp and hold for 1 h | Above measuring limit | ~0.5 | Correct phase | Powdery sample detaches from quartz. |
| SnSe$_2$ drop-cast on quartz | @ 400 °C with 0.1 °C/s ramp and hold for 6 h | Above measuring limit | ~0.5 | Correct phase | Powdery sample detaches from quartz. Turns white after annealing |
| SnSe$_2$ drop-cast on quartz | @ 500 °C with 0.1 °C/s ramp and hold for 1 h | ~ 170 MOhm | ~1 | Correct phase | Powdery sample detaches from quartz. Turns white after annealing |
| SnSe$_2$ drop-cast on quartz | @ 600 °C with 0.1 °C/s ramp and hold for 1 h | ~ 0.070 MOhm | >3 | Correct phase | Powdery sample detaches from quartz. Turns white after annealing |

### 3.3. Investigation of the role of composition of the starting inks in drop-cast samples

Regarding the composition of the starting inks, attention was paid to non-toxicity of solvents and additives. Samples were prepared from starting inks having either (i) just ethanol as solvent, or (ii) ethanol as solvent and terpineol as additive to improve rheological properties in view of ink-jet printing. In presence or absence of terpineol, any effect was observed neither on the phase of the final assemblies inspected by Raman spectroscopy, nor on the conductance of the final assemblies. Hence it was concluded that the addition of terpineol for ink-jet printing had no drawbacks.

## 3.4. Investigation of nanoflake alignment in drop-cast samples

Due to strong structural and electronic anisotropy of layered Van der Waals TMDs, a good alignment of deposited nanoflakes is highly desirable to improve inter-flake transport and thus macroscopic properties.
In order to check nanoflake alignment, the X-rays rocking curve was measured in a not annealed $WSe_2$ drop-cast sample. A very broad (∼ 30°) rocking curve was found, indicating random orientation of the deposited nanoflakes. Also X-rays θ-2θ patterns of $WSe_2$ (Fig. 3, left) and $SnSe_2$ (Fig. 4, left) drop-cast samples exhibited all Bragg reflections, confirming the random orientation of the nanoflakes. Ideally maximal nanoflake connectivity would be obtained in adjacent flakes touching each other by lateral sides. For this reason, uniaxial pressing was carried out on drop-cast samples, in order to improve average nanoflake alignment, as well as packing.

## 3.5. Uniaxial pressing of drop-cast samples

In order to improve nanoflake alignment and packing, thus improving inter-flake conductance, an additional step of uniaxial pressing was introduced in the preparation protocol.
Drop-cast samples were pressed at 1 GPa for 1 h. Polyester and kapton were chosen as flexible substrates that endure pressing without shattering. The effectiveness of pressing on $WSe_2$ nano-flake assemblies was apparent by optical imaging. Indeed, after pressing the optical reflectivity increased significantly, as shown in Fig. 5 (left). In addition, the surface roughness inspected by SEM imaging became smoother and more uniform after pressing, as shown in Fig. 5, where images before (middle) and after (right) pressing are compared.

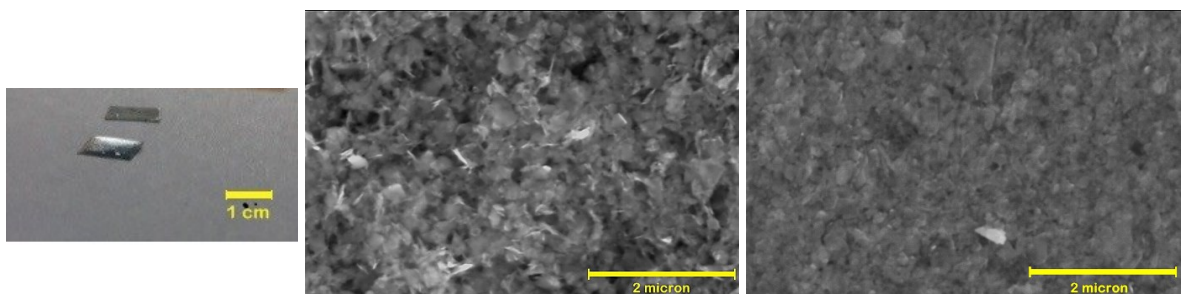

**Fig. 5:** Left: picture of $WSe_2$ drop cast samples before (upper sample) and after (lower sample) pressing, showing increased reflectivity in the latter case. Middle and right: comparison of SEM Secondary Electron images of $WSe_2$ drop cast samples taken before (left) and after (right) pressing, showing improved surface smoothness and uniformity in the latter case (right).

$SnSe_2$ nanoflake assemblies, suffered of powdery texture, which made it challenging to carry out any processing and measurement. To tackle this problem, a sandwich architecture with kapton foils was developed, as sketched in Fig. 6. $SnSe_2$ was successively (1) drop cast on a kapton foil, (2) annealed at 200°C for 1 h in UHV conditions, (3) sandwiched between other two kapton foils, (4) pressed at 1 GPa for 1 h. The kapton foils were never removed to avoid cracking and electrical contacts were obtained by punching holes through the kapton foils.

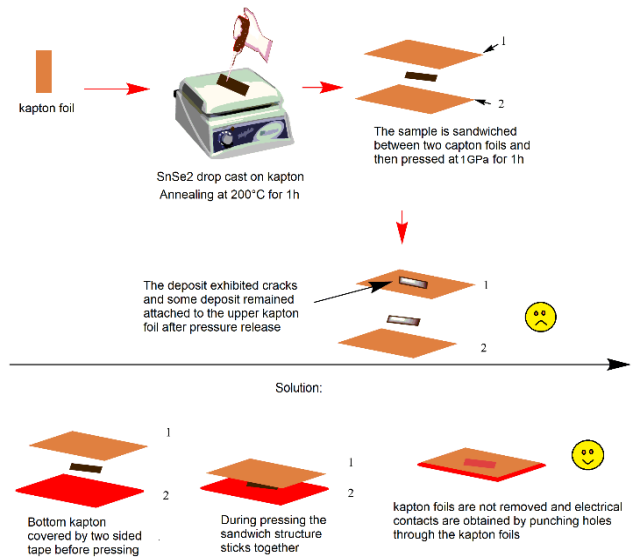

**Fig. 6:** Sketch of the preparation protocol of SnSe$_2$ nanoflake assemblies, developed to tackle the powdery texture of these samples.

From SEM imaging, it appeared that surface roughness of SnSe$_2$ nanoflake assemblies became smoother and more uniform after pressing and packing of nanoflakes improved (Figs. 7 and 8 for not annealed and annealed samples, respectively). The thickness of the deposits was significantly reduced by a factor 2 or more after pressing (right panels of Figs. 7 and 8 for not annealed and annealed samples, respectively).

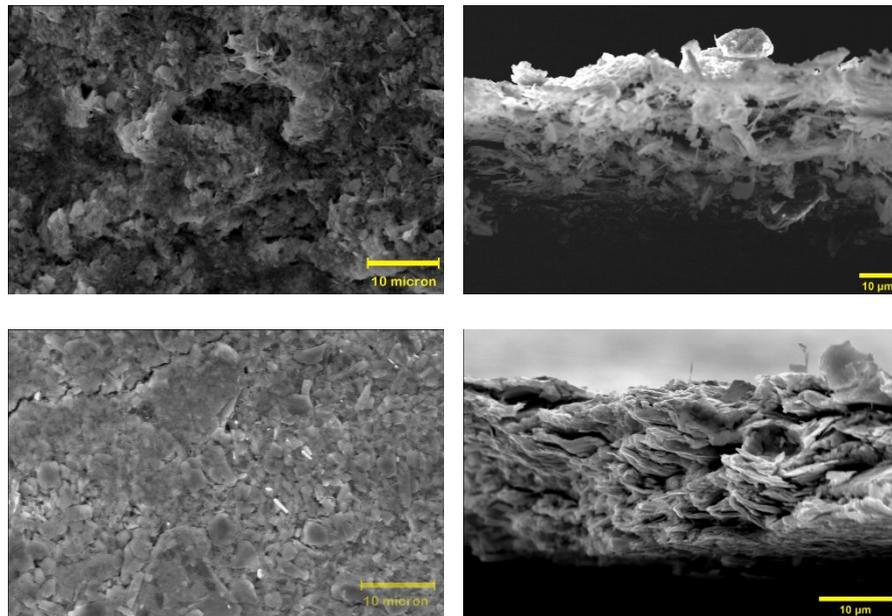

**Fig. 7:** SEM Secondary Electron images showing surface smoothness and uniformity of SnSe$_2$ drop cast assemblies, not annealed, before (top) and after (bottom) pressing at 1 GPa for 1 h. The left images show a top view, while the right images show a cross section, where it is seen that the thickness was nearly halved from ≈40-50 μm to ≈20 μm after pressing.

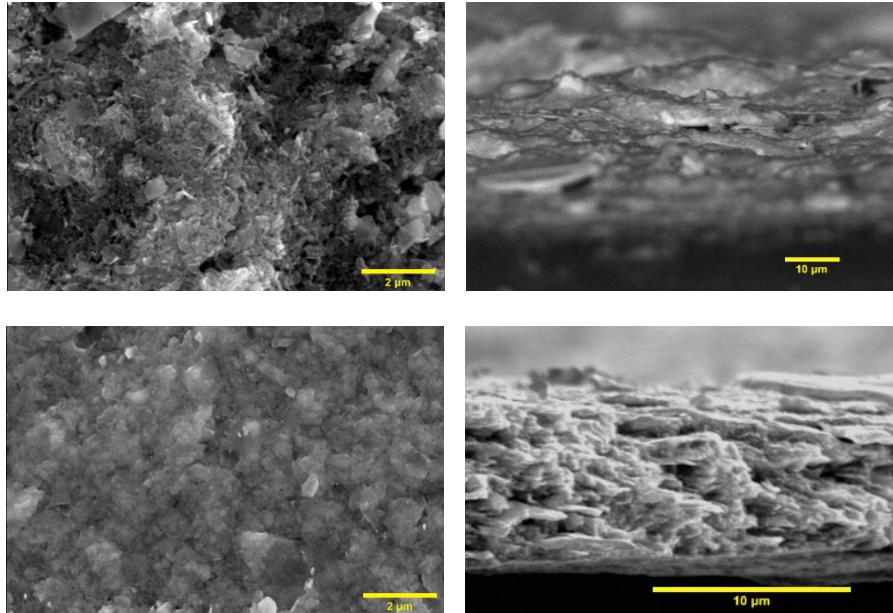

**Fig. 8:** SEM Secondary Electron images showing surface smoothness and uniformity of SnSe$_2$ drop cast assemblies, annealed at 200°C, before (top) and after (bottom) pressing at 1 GPa for 1 h. The left images show a top view, while the right images show a cross section, where it is seen that the thickness was reduced from ≈30 μm to ≈10 μm after pressing.

The effectiveness of pressing in aligning nano-flake assemblies drop-cast on Kapton was apparent in X-rays θ-2θ patterns (Fig. 9), particularly in the case of SnSe$_2$, where 00l peaks were clearly dominant in pressed samples, indicating 00l preferential orientation, oppositely to what observed in not pressed samples (Fig. 3 left, and Fig. 4 left).

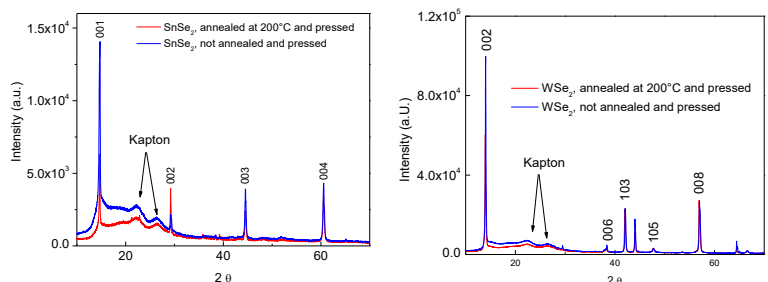

**Fig. 9:** X-rays patterns of pressed SnSe$_2$ (left) and WSe$_2$ (right) nanoflake assemblies drop-cast on kapton, either annealed or not annealed.

### 3.6. Electric and thermoelectric transport characterization of drop-cast samples

Resistance versus temperature curves of WSe$_2$ drop cast samples showed steep rise with decreasing temperature and high magnitudes, around 10 MOhm at room temperature. This is consistent with the few literature results on electrical transport in nanoflake assemblies of 2H phase TMDs obtained by LPE and either drop-casting or ink-jet printing, indeed resistances of tens of MOhms or larger were obtained in MoS$_2$ and MoSe$_2$ nanoflake assemblies [12,14,15,19]. In Fig. 10 (left), resistance curves of WSe$_2$ assemblies drop cast on quartz, annealed at different temperature 200°C-300°C and for different times 6-36 h, are shown. For annealing temperatures below 400°C, the correct stoichiometry was preserved, hence the decrease of resistance observed for higher annealing temperature (300°C versus 200°C) and for longer annealing times (36 h versus 6h) can be ascribed to increased sintering and better inter-flake connectivity. The thermopower measured on the lowest resistance sample annealed at 300°C, was negative (Fig. 10, right). For comparison, literature data of thermopower on WSe$_2$ single flakes show large variations in magnitude and sign as a function of gate voltage [37].

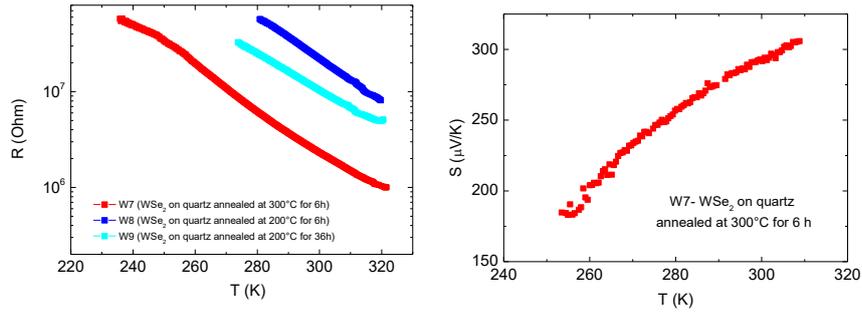

**Fig. 10:** Left: four-probe resistance of the WSe$_2$ drop-cast assemblies annealed at 200°C – 300°C for 6-36 h times. Right: thermopower of a WSe$_2$ drop-cast assembly annealed at 300°C for 6 h.

Resistance measurements carried out on WSe$_2$ drop-cast samples obtained from terpineol-free and terpineol added inks showed no appreciable differences, indicating that possible terpineol residues at inter-flake boundaries were not an issue.

Resistance was also measured in WSe$_2$ and SnSe$_2$ drop-cast assemblies, deposited on kapton and pressed. The additional pressing step in the preparation gave much lower resistance values for both compositions. In the case of WSe$_2$ samples, pressed and either annealed at 200°C or not annealed, the lowest room temperature resistance was ~90 kOhm in one annealed sample (Fig. 11, left). In the same WSe$_2$ samples, thermopower measurements gave very large values of the Seebeck coefficient, namely 2.5 mV/K and 1.8 mV/K in the annealed and not annealed samples at room temperature, respectively (Fig. 11, middle). The calculation of the thermoelectric power factors S$^2\sigma$ requires extracting the conductivity $\sigma$ from the measured resistance. This can be done only approximatively, using the second Ohm's law, which is valid for an ideally compact and dense sample, where the current flow is uniform across the sample. On the other hand, for a nanoflake assembly with some porosity and non-ideal inter-flake conductance, the current paths are meandering at the microscopic scale and a correction factor of the order of few units should be introduced to account for the actual longer current path and reduced current cross section. Given that any estimation of such correction factor would be arbitrary, it was set to unity. This choice may yield an underestimation of conductivity, similar for all the samples, and thus it represents a lower limit for their power factors, giving a room temperature power factor 3.0x10$^{-4}$ mW m$^{-1}$ K$^{-2}$ in the annealed sample and 1x10$^{-5}$ mW m$^{-1}$ K$^{-2}$ in the not annealed sample (Fig. 11, right).

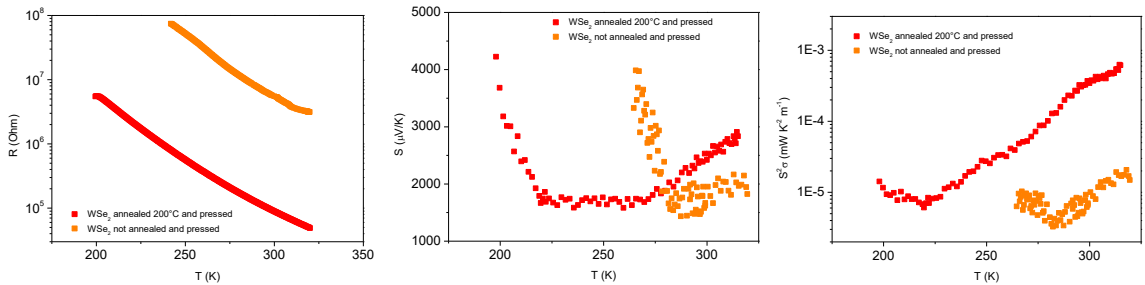

**Fig. 11:** Resistances (left), Seebeck coefficients (middle), and power factors (right) of WSe$_2$ assemblies drop cast on kapton, pressed at 1 GPa for 1 h, and either annealed at 200°C or not annealed.

In the case of pressed SnSe$_2$ drop-cast samples, room temperature resistance was found to be 250 kOhm in one annealed sample and as low as 2.1 kOhm in one not annealed sample (Fig. 12, left), in both cases much lower than the room temperature resistances found in WSe$_2$ assemblies, possibly due to the better flake alignment induced by pressing (see Fig. 9), as well as lower intra-flake conductivity of SnSe$_2$ with respect to WSe$_2$. Notably, in the case of SnSe$_2$, resistance was lower in the not annealed samples than in the annealed ones, oppositely to the case of WSe$_2$. Our interpretation of this finding is that annealing improved sintering (and thus it decreased the resistance), but at the same time it increased the brittleness of the deposit (and thus it might increase the resistance of the assembly, if microcracks occurred, during processing or handling of the sample on the flexible substrate). The former effect could be dominant for the WSe$_2$ drop-cast samples, resulting in a net beneficial effect of annealing, while the latter could be dominant for the SnSe$_2$ drop-cast samples, which were powderier, resulting in a net detrimental effect of annealing. In one not annealed sample, resistance was significantly lower than in all the other samples; this fact was not related simply to larger thickness, because the thickness was only a factor ~2 larger than that of the other samples, but it should be rather related to higher packing on the nanoflakes. In this not annealed sample with the lowest resistance, thermopower measurements (Fig. 12, middle) indicated a room temperature Seebeck coefficient S around -100 μV/K, while on the annealed sample with 2.1 MOhm room temperature resistance, the Seebeck coefficient was -700 μV/K. The thermoelectric power factors S$^2$σ at room temperature turned out to be 2.2x10$^{-4}$ mW m$^{-1}$ K$^{-2}$ in the not annealed sample with the lowest resistance and 2x10$^{-5}$ mW m$^{-1}$ K$^{-2}$ in the annealed sample with 2.1 MOhm room temperature resistance (Fig. 12, right). On the whole, a significant scattering of electric and thermoelectric properties was observed between TMD samples obtained by the same procedure. Indeed, the values of resistance, determined mostly by the inter-flake contribution, varied by more than one order of magnitude from sample to sample. On the other hand, the values of the Seebeck coefficient varied within a smaller range, as they are rather determined by intra-flake electronic properties such as carrier density; however, carrier density in TMDs is sensitively dependent on doping by adatoms and impurities at the surface, and thus it may also vary appreciably from sample to sample. As a result, the thermoelectric power factors, dependent on conductivity and squared Seebeck coefficient, exhibited large sample-to-sample variations, associated to variations of both conductivity and Seebeck coefficient, but mostly to variations of conductivity. This fact indicates that full control of fabrication was not yet achieved, likely due to the crucial role of microstructure and nanoflake packing. Criticality of sample handling might be also an issue for the measured macroscopic resistance.

The obtained values of power factors of WSe$_2$ and SnSe$_2$ pressed nanoflake assembles are small for applications, the low magnitude mainly ascribed to the high resistance values; however, they are 2-3 orders of magnitude larger in comparison with the values found for not pressed assemblies, indicating an edge of improvement in terms of optimized and fully-controlled preparation procedure.

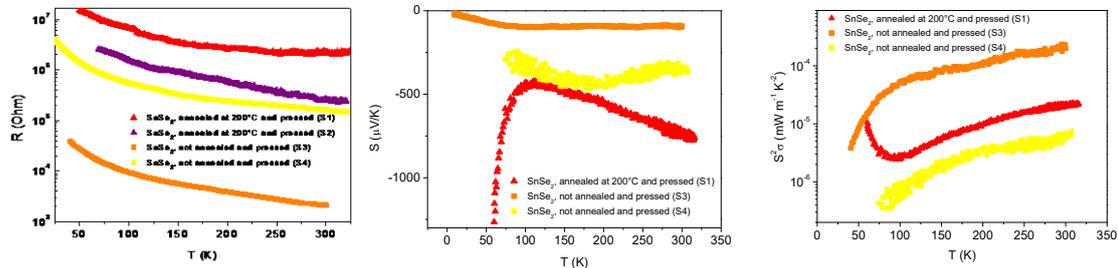

**Figure 12:** Resistances (left), Seebeck coefficients (middle), and power factors (right) of SnSe$_2$ samples drop cast on kapton, pressed at 1 GPa for 1 h, and either annealed at 200°C or not annealed.

### 3.7. Electric transport characterization of WSe$_2$ ink-jet printed samples

In ink-jet printed samples, resistances of higher magnitude and steeper temperature behavior were obtained, even after uniaxial pressing. Conductive ink-jet printed samples with typical lateral dimensions of the order of millimeters were obtained only with the WSe$_2$ composition, because SnSe$_2$ flakes easily clogged the ink-jet printer nozzles. In Fig. 13 we show resistance measurements on annealed and not annealed ink-jet printed WSe$_2$ samples on Kapton substrates. Room temperature values of few MOhm were obtained and steep temperature increases above 100 MOhm within few temperature decades were measured. Note that without the pressing step, no electrical conduction at all was detected. We do not comment of the difference between annealed and not annealed ink-jet printed WSe$_2$ samples, as with such large resistances this difference is comparable to the observed sample-to-sample difference between samples prepared in the same conditions.

In order to explain the different conductance of drop cast and ink-jet printed assemblies, thorough microstructural analysis was carried out on WSe$_2$ samples, pressed on kapton. No significant differences were found between drop cast and ink-jet printed samples, except that in drop-cast samples some larger flakes with micrometric size were found, while in ink-jet printed samples the flake size was generally sub-micrometric (see supplementary material for SEM images, Fig.s S2-S6). However, from SEM cross-sectional images, it appeared that in ink-jet printed pressed samples the thickness was more uniform than in drop-cast pressed samples, and, most importantly, that the amount of deposited material was significantly different in the two cases (see supplementary material for SEM cross-sectional images, Fig. S6). Hence, the different conductance of drop cast and ink-jet printed samples must be ascribed to the different amount of deposited material, namely a factor 5-10 difference estimated from SEM cross-sectional images, which just corresponds to a similar factor difference in resistances (compare Fig.s 11 (left) and 13). Indeed, assembling of the nanoflakes critically depends on the fluid dynamics and related drag forces on the nanoflakes themselves, which are different for the case of picoliter ink-jetted drops and drop cast ones. Furthermore, the overprinting of drops further reorganizes the already-printed layers, making the overall transport properties not obviously scalable with the number of layers.

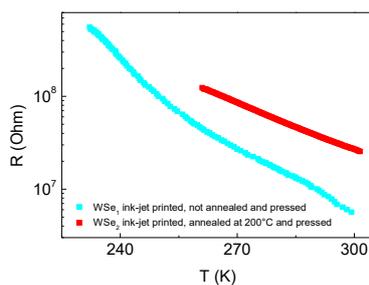

**Figure 13:** Resistance measurements of $WSe_2$ samples ink-jet printed on kapton, pressed at 1 GPa for 1 h, and either annealed at 200°C or not annealed.

## 4. Conclusions

In this work, 2H phase $SnSe_2$ and $WSe_2$ nanoflakes were obtained by LPE and deposited by ink-jet printing and drop-casting on different substrates, namely mica, kapton and quartz. In the case of drop-casting, optimal fabrication parameters with respect to amount of deposited material, homogeneity and electrical conductivity of the deposits were identified in terms of multiple deposition steps and annealing temperature and time. In the case of ink-jet printing, fabrication was investigated in terms of drop spacing, spacing between successive adjacent lines, number of stacked layers, and time interval between successive layers. The systematic investigation of fabrication conditions evidenced that annealing at 400°C improved sintering while preserving the nominal stoichiometry, thus resulting in samples with measured resistances in the MOhm range at room temperature. A successive step of uniaxial pressing improved the nanoflake packing further, resulting for $SnSe_2$ samples in measured room temperature resistances in the kOhm range, Seebeck coefficient values up to 700 μV/K and thermoelectric power factors $S^2\sigma$ up to $2.2 \times 10^{-4}$ mW m$^{-1}$ K$^{-2}$ and for $WSe_2$ samples in measured room temperature resistances in the 100 kOhm range, Seebeck coefficient values up to 2.5 mV/K and thermoelectric power factors $S^2\sigma$ up to $3.0 \times 10^{-4}$ mW m$^{-1}$ K$^{-2}$. Conductive $WSe_2$ patterns were obtained also by ink-jet printing and uniaxial pressing, with room temperature resistance values of few MOhm.

In perspective, the development of simple, scalable and cost-effective manufacturing techniques to prepare Van der Waals compounds on flexible substrates with good thermoelectric properties offers interesting opportunities in the emerging field of wearable energy harvesters, as components in self-powered healthcare devices.

## Acknowledgments
We thank Francesco Bonaccorso, Maurizio Vignolo, and Emilio Bellingeri for useful discussions. We acknowledge financial funding from the JTC 2017 FlagERA project MELODICA.

# Supplementary material

**Ink-jet printing and drop-casting deposition of 2H-phase SnSe$_2$ and WSe$_2$ nanoflake assemblies for thermoelectric applications**

B. Patil [1], C. Bernini [1], D. Marré [2,1], L. Pellegrino [1], I. Pallecchi [1]

[1] *CNR-SPIN, Corso Perrone 24, 16152 Genova, Italy,*
[2] *Università di Genova, Dipartimento di Fisica, Via Dodecaneso 33, 16146 Genova, Italy*

## 1. Ink-jet printing technical details

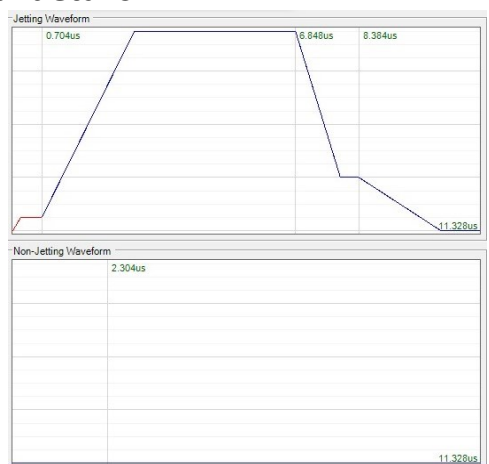

**Figure S1:** Jetting and non-jetting waveform of the voltage driving the piezo pumps of the ink-jet printer nozzles for printing WSe$_2$ nanoflake inks in ethanol solution. Dimatix DMP-2831 Fujifilm inkjet printer was used.

## 2. Microstructural characterization of ink-jet printed and drop-cast and pressed WSe$_2$ on kapton

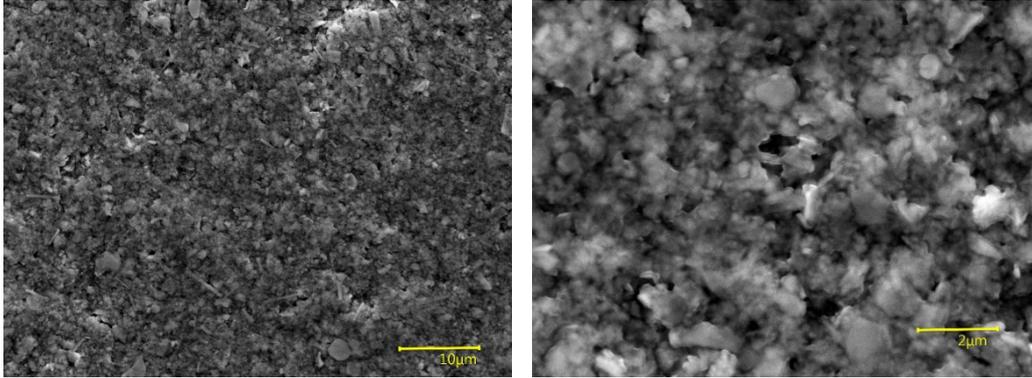

**Figure S2:** WSe$_2$ ink-jet printed on kapton, not annealed and pressed at 1 GPa for 1 h.

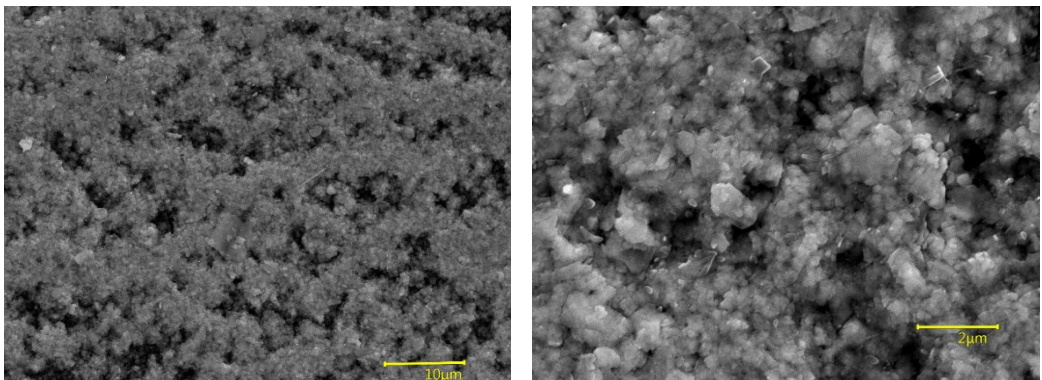

**Figure S3:** WSe$_2$ ink-jet printed on kapton, annealed at 200°C and pressed at 1 GPa for 1 h.

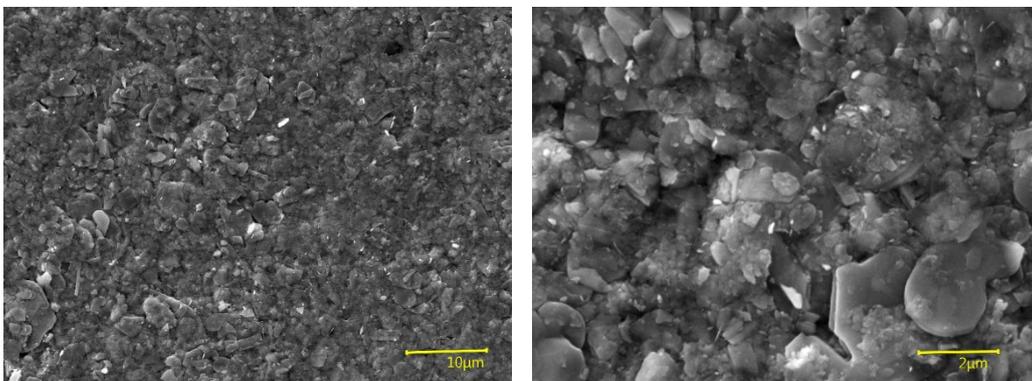

**Figure S4:** WSe$_2$ drop-cast on kapton, not annealed and pressed at 1 GPa for 1 h.

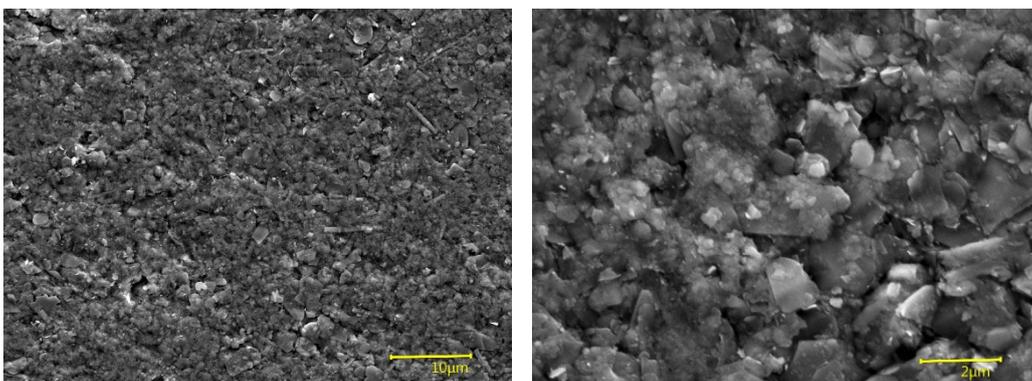

**Figure S5:** WSe$_2$ drop-cast on kapton, annealed at 200°C and pressed at 1 GPa for 1 h.

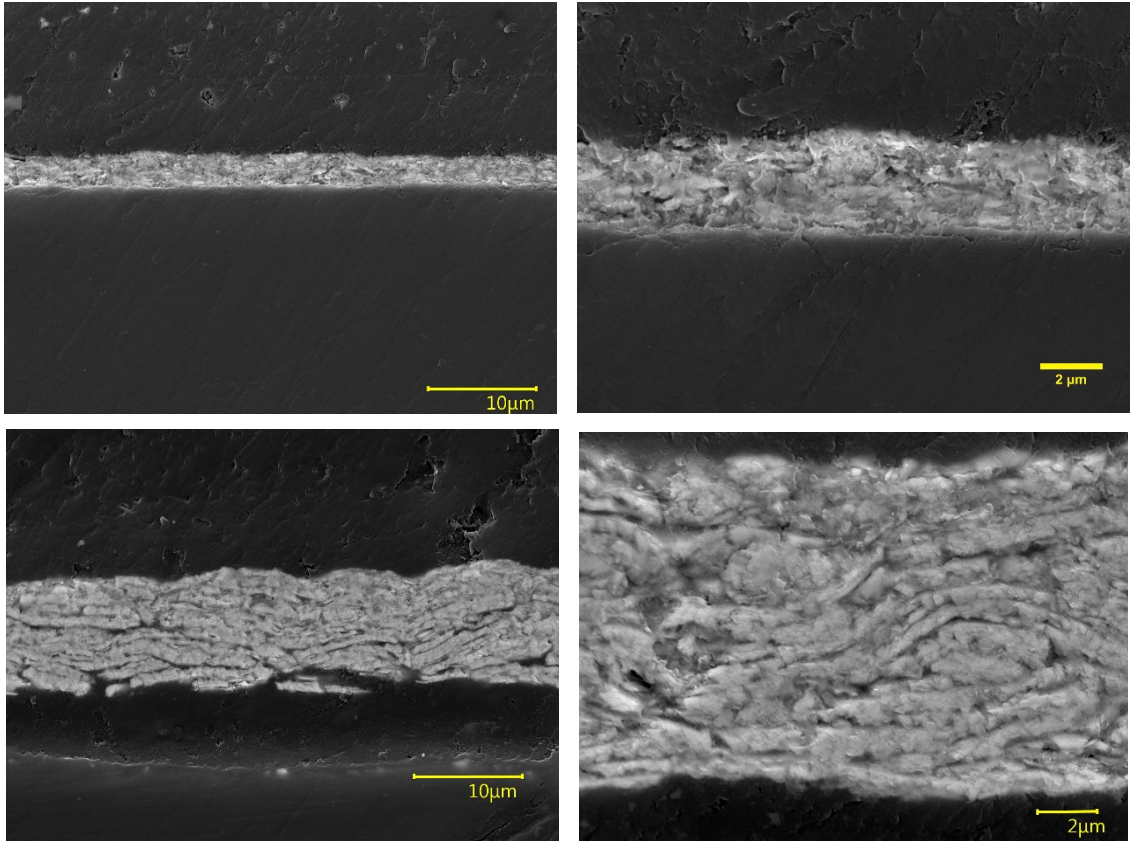

**Figure S6:** WSe$_2$ ink-jet printed (images in the top panels) and drop-cast (images in the bottom panels) on kapton, not annealed and pressed at 1 GPa for 1 h. A factor 5-10 difference in the amount of deposited material is seen.